\documentstyle[12pt,epsfig]{article}
\def \sen{\mathop{\rm sen}\nolimits}
\def \cos{\mathop{\rm cos}\nolimits}

\let \nm=\nonumber
\let \n=\noindent

\let \q=\qquad
\let \o=\over

\newcommand{\beqn}{\begin{eqnarray*}}
\newcommand{\eeqn}{\end{eqnarray*}}
\newcommand{\beq}{\begin{eqnarray}}
\newcommand{\eeq}{\end{eqnarray}}
\newcommand{\eqn}{\begin{equation}}
\newcommand{\een}{\end{equation}}
\newcommand{\ba}{\begin{array}}
\newcommand{\ea}{\end{array}}

\begin{document}

 \date{\ }
\title{${\rm QED}_{2+1}$: the Compton effect.}
\author{ J. Mateos Guilarte \& M. de la Torre Mayado\\
 Departamento de F\'{\i}sica. Facultad de Ciencias\\  Universidad de Salamanca.
Salamanca 37008\\ SPAIN}
\maketitle

\begin{abstract}

The Compton effect in a two-dimensional world is compared with the same
process in
ordinary three-dimensional space.

\end{abstract}

\section{Introduction}

QED, the strange theory of light and matter, \cite{R.Feynman85}, embraces in a
unified dogma several of the most successful physical doctrines. Quantum
electrodynamics describes the interactions of the electromagnetic field with
electrons and positrons in a framework ruled by the laws of special relativity
and quantum mechanics. According to Feynman, see Reference
\cite{R.Feynman85}, QED
is the jewel of physics: there is no significant difference between experiment
and theory. Nevertheless, nobody understand why Nature works that way, quoting
Feynman again. The lack of comprehension is partly due to the conceptual
difficulties of both special relativity and quantum mechanics, so far from
common sense, and partly due to the complexity of the phenomena involved. There
are too many degrees of freedom entering the game and a nightmare of divergences
must be tamed by proper physical insight.

During the eighties, interesting investigations were devoted to QED in a
space-time of (2+1)-dimensions, see e.g. Reference \cite{S.Deser82}. The
research was pushed forward by either purely theoretical reasons, to study
conventional QED at the infinite temperature limit, or condensed matter
experiments: both the Quantum Hall Effect \cite{R.Prange90} and High $T_c$
Superconductivity \cite{J.C.Phillips} are many-body quantum phenomena including
interactions of charged fermions with the electromagnetic field that essentially
occur in two-dimensions.

Perturbation theory of planar quantum electrodynamics is rich enough to be
compared with perturbative features of the theory of photons, electrons and
positrons in three - dimensional space. The theoretical analysis of
photon-electron scattering performed within the framework of ${\rm
QED}_{2+1}$ allows for a
comparative study with the same process in three dimensions. Besides the
academic
interest, we find it fruitful to enlarge the list of lowest order processes
analyzed in QED.
We compute the differential cross section, a length in two-dimensions
\cite{Lapidus82}, up to second order in perturbation theory starting from the usual
plane wave expansion of field operators. We then compare the results in
different
physical situations with the outcome of the very well known analysis in
three dimensions.

The organization of the paper is as follows: in Section \S 2 we briefly
present ${\rm QED}_{2+1}$
Perturbation Theory and its application to compute $S$-matrix elements.
Section \S 3 is divided into two
sub-sections: calculation of both the differential and total {\it
cross-lengths} of scattering for the Compton
effect is performed in \S {3.1}. The planar analogue of the Thomson and
Klein-Nishina formulas are
discussed in \S {3.2} and compared with the behaviour of the same
expressions in three dimensions.
Finally, in the Appendices the definition of the Dirac and electromagnetic
fields in three-dimensional
Minkowski space is given. Also, some useful formulas are collected and the
conventions to be used throughout the paper are fixed.

\section{Quantum electrodynamics in the plane.}

Quantum electrodynamics in the plane, ${\rm QED}_{2+1}$, describes the
interaction of two-dimensional electrons, positrons and photons by means
of the quantum field theory derived from the Lagrangian density
\eqn
{\cal L} = { \cal L}_0 + {\cal L}_I \label{1}
\een
with the free-field Lagrangian density
\eqn
{\cal L}_0 = N \left[  c  { \bar { \psi } }(x) \left(  i \hbar  \gamma^{\mu}
\partial_{\mu} - m c \right)  \psi(x)  - { 1\o 4} f_{\mu \nu}(x) f^{\mu \nu}(x)
\right]  \label{2}
\een
where: (1) $\psi(x)$ and ${ \bar { \psi } }(x) = \psi^{\dag}(x) \gamma^0 $
are the
Dirac fields that have relativistic matter particles of spin $1 \o 2$,
electrons and positrons, as quanta. We take the charge of the electron as
$ q = - e < 0$, and $m$ is the mass of the particle. Study of the Dirac
free-field in (2+1) dimensions and its quantization can be found in
Appendix \ref{apdC}. (2) $
a_{\mu}(x) $, $\mu = 0, 1, 2$, is the three-vector electromagnetic
potential and $f_{\mu \nu}$ the
associated antisymmetric tensor to the electromagnetic field :
\[
f_{\mu \nu} = \partial_{\mu} a_{\nu} - \partial_{\nu} a_{\mu} .
\]

A Lorentz-covariant formulation of the free electromagnetic field in the
plane, the quantization
procedure leading to the identification of polarized photons as its quanta,
is also developed in the Appendix \ref{apdD}.
(3) $\hbar $ and c are respectively the Planck constant and the speed of
light in vacuum.
The interaction Lagrangian density is
\eqn
{\cal L}_I = N\left[ e { \bar { \psi } } (x) \gamma^{\mu } a_{\mu }(x) \psi(x)
\right]  \equiv N \left[ - {1 \o c} j^{\mu}(x) a_{\mu}(x) \right]  \label{3}
\een
which couples the conserved current $j^{\mu}(x) = (- e) c { \bar { \psi } } (x)
\gamma^{\mu } \psi(x) \equiv ( c \rho(x), {\vec j}(x) )$ to the electromagnetic
field. We have defined the Lagrangian density as  a normal product, $N [\
]$, each creation operator
standing to the left of any annihilation operator, to
ensure that the vacuum expectation values of all observables vanish, (we
follow the conventions of \cite{Mandl}, chapters 4,5,6 and 7 ).

The action integral $S$ for quantum electrodynamics in three-dimensional
space-time is therefore
\eqn
 S = \int d^3 x \  N \left[ c  { \bar { \psi } }(x) \left( \gamma^{\mu} \left( i
\hbar \partial_{\mu} + { e \o c} a_{\mu}(x) \right) - m c \right)  \psi(x)
 - { 1\o 4} f_{\mu \nu}(x) f^{\mu \nu}(x) \right]   \label{4}
\een

In relativistic quantum field theory it is convenient to work in
natural units (n.u.), $\hbar = c =1$. In these units the fundamental dimensions
are the mass (M), the action (A) and the velocity (V) instead of the mass (M),
length (L) and time (T) that are the fundamental dimensions in c.g.s. (or
S.I.) units. It is interesting to analyse the dimensions of the free-fields and
constants that appear in the action $S$ for ${\rm QED}_{2+1}$, and to
compare them with
the dimensional features of the same magnitudes in ${\rm QED}_{3+1}$. In
general, given
the dimension of the action integral,
\eqn
 S= \int d^{d+1} x \ \cal L ,
\een
where $\cal L$ is defined by (\ref{2}) and (\ref{3}); the dimension of the
fields is determined
from the kinetic terms, and then the dimension of the coupling constants is
fixed. We find that:

\begin{center}
\begin{tabular}{|l|c|c|}
\hline & &  \\  Quantity & c.g.s. & n.u. \\ & &  \\
\hline & &  \\ Action $S$ & $ {\rm M} {\rm L}^2 {\rm T}^{-1}$  & $1$  \\ & & \\
Lagrangian density ${\cal L}$ & $ {\rm M} {\rm L}^{2-d} {\rm T}^{-2}$  &
${\rm M}^{d+1} $ \\ & &\\ Electromagnetic field $a_{\mu}(x)$ & $ {\rm M}^{1\over
2} {\rm L}^{2-{d\over 2}} {\rm T}^{-1}$  & ${\rm M}^{{d-1}\over 2}$ \\ &
&\\ Dirac
fields $\psi(x)$ and $\bar{\psi}(x)$  & $  {\rm L}^{-{d\over 2 }}$ &
${\rm M}^{d \over 2}$ \\ & &\\ Electric charge $e$ & $ {\rm M}^{1\over 2} {\rm
L}^{d\over 2} {\rm T}^{-1} $ &
$ {\rm M}^{{3-d} \over 2}$ \\ & &\\
 Mass of the electron $m$ & $ {\rm M}$  & ${\rm M}$  \\ & & \\
\hline
\end{tabular}
\end{center}

The c.g.s. dimensions of $e^2$ in $d=3$ are $[ e^2] = {\rm M} {\rm L}^3 {\rm
T}^{-2} \equiv [ \hbar c ]$ due to the fact that the Coulomb force
decreases as ${1\over r^2}$. In natural units, however, the electron charge
is dimensionless. The fine structure constant $\alpha \approx {1\over
137.04}$ is given by,
\eqn
 \alpha = {e^2 \over 4 \pi \hbar c} \q {\rm (c.g.s.)}\q {\rm or} \q \alpha
= {e^2 \over 4 \pi }\q {\rm (n.u.)}
\label{5}
\een
This means that one can take $\alpha$ or the electron charge as a good
expansion parameter because both of them are dimensionless in n.u..

Things are different in a two-dimensional world: if $d=2$, the electron
charge is not dimensionless but $[e^2] =
{\rm M} {\rm L}^2 {\rm T}^{-2}$ in c.g.s. units, or $[e^2]= {\rm M}$ in
n.u. because the Coulomb force is
proportional to ${1\over r}$. The fine structure constant is still the
expansion parameter for a perturbative
treatment of
${\rm QED}_{2+1}$ but we must keep in mind that the electron charge has
dimensions. The dimension of the product
of $e^2$ times the Compton wave length ${\hbar \over m c}$ is $[e^2  {\hbar
\over m c}] = {\rm M L^3 T^{-2}}$ or
$[{e^2 \over m}] = {1}$, respectively in c.g.s. or n.u. systems. Therefore,
we express the fine structure constant
as:
\eqn
 {\alpha} = { e^2 \over 4 \pi m c^2} \q {\rm (c.g.s.)} \q {\rm or} \q
{\alpha} = { e^2 \over 4 \pi m } \q {\rm
(n.u.)}
\label{6}
\een
bearing in mind that $e^2$ is not dimensionless in natural units when
$d=2$. Thus, in (2+1)-dimensional Minkowski
space the fine structure constant is $(electron\  charge)^2 \over electron\
mass$, (up to $4 \pi$ factors).

Perhaps a rapid comparison of the several systems of units used in
electromagnetism will help to clarify this
point: in the above formulas we have adopted the rationalized
Lorentz-Heaviside system of electromagnetic units;
that is, the $4\pi$ factors appear in the force equations rather than in
the Maxwell equations, and the vacuum
dielectric constant $\epsilon_0$ is set equal to unity \cite{Jackson}. In rationalized mks units,
the fine structure constant is defined as:
\eqn
\alpha = {e^2 \over 4 \pi \epsilon_0
\hbar c} \ (d=3) \q {\rm or} \q {\alpha} = { e^2 \over 4 \pi a_0 m c^2} \
(d=2) \label{7}
\een
where $a_0$ has dimensions of permitivity by length. We define $a_0 =
\epsilon_0 {\hbar \over m c}$, the
permitivity of vacuum times the fundamental length of the system. Then,
\eqn
 {\alpha} =  {e^2 \over 4 \pi \epsilon_0
\hbar c}\equiv  { e^2 \over 4 \pi a_0 m c^2}  \approx {1 \over 137.04}
\label{8}
\een
where the rationalized electric charges ${e^2 \over \epsilon_0}$ and ${e^2
\over a_0}$ have different dimensions.

 The Hamiltonian $H$ of the system splits into the free $H_0$ and the
interaction
 $H_I$ Hamiltonians. $H_I$ can be treated as a perturbation since the
dimensionless
coupling constant, characterizing the photon-electron interaction in the plane,
is small enough: $ {\alpha} \approx {1 \over 137.04}$. In the interaction
picture the S-matrix expansion is
\eqn
S = \sum_{n=0}^{\infty} { (-i)^n \over n!} \int \cdots \int d^3 x_1 d^3 x_2
\cdots d^3 x_n \ T \left\{ {\cal H}_I(x_1){\cal H}_I(x_2) \cdots {\cal H}_I(x_n)
\right\}  \label{9}
\een
where $T\{ \} $ is the time-ordered product. The ${\rm QED}_{2+1}$ interaction
Hamiltonian density
\eqn
{\cal H}_I(x) = - {\cal L}_I (x) = - e N\left[ { \bar { \psi } } (x) \gamma^{\mu
} a_{\mu }(x) \psi(x)
\right]  \label{10}
\een
determines the basic vertex part of the theory.

For the $|i \rangle \rightarrow |f
\rangle $ transition, the S-matrix element is given by
\eqn
\langle f | S| i \rangle = \delta _{fi} + \left[ (2\pi )^3 \delta^{(3)} ( P_f -
P_i) \prod_{\rm ext.} \left( {m \over A E} \right) ^{1/2} \prod_{\rm ext.}
\left( {1 \over 2 A \omega} \right) ^{1/2} \right] {\cal M} . \label{11}
\een
Here, $P_i$ and $P_f$ are the total three-momenta of the initial and final
states, the products extend over all external particles, and $A = L^2$ is a
large but finite area in the plane. $E$ and $\omega$ are the energies of the
individual external fermions and photons, respectively. ${\cal M}$ is the
Feynman
amplitude such that: ${\cal M} ={ \sum_{n=1}^{\infty}} {\cal M}^{(n)}$, and the
contribution to ${\cal M}^{(n)}$ (nth order in perturbation theory) from each
topologically different graph is obtained from the Feynman rules
\cite{Mandl}. We only
enumerate the fundamental differences with respect to the Feynman rules in ${\rm
QED}_{3+1}$:
\begin{itemize}
\item  The four-momenta of the particles are now three-momenta.
\item  For each initial and final electron or positron there is only one
label, $s = 1$ , that characterizes the spin state.
\item  For each initial and final photon there is also only one label, $r = 1$,
that characterizes the polarization state.
\item  Initial and final electrons and positrons have associated
two-component spinors.
\item The $\gamma$-matrices arising at vertices and the $S_F$-functions,
coming from propagation of
 internal fermion lines, are $2 \times 2$ matrices.
\item For each three-momentum $q$ which is not fixed by energy-momentum
conservation one must  carry out the integration $ (2 \pi )^{-3} \int d^3 q $.
\end{itemize}

\section{A ${\rm QED}_{2+1}$ Lowest Order Process:    The Compton \break Effect}

In this Section we shall discuss the scattering cross-section, a {\it
cross-length} in d=2, for planar Compton
scattering up to second order in perturbation theory.

\subsection{ Compton Scattering }

 The S-matrix element for the transition
\eqn
|i\rangle = c^{\dag}(\vec p) b^{\dag}(\vec k) |0\rangle \longrightarrow
|f\rangle =
c^{\dag}(\vec {p'}) b^{\dag}(\vec{k'}) |0 \rangle \label{12}
\een
to second order in $e$ is:
\eqn
S^{(2)} = - e^2 \int d^3 x_1 d^3 x_2 N \left[   {\bar \psi}(x_1) \gamma^{\alpha}
a_{\alpha} (x_1) i S_F(x_1-x_2) \gamma^{\beta} a_{\beta}(x_2) \psi(x_2)
\right] =
S_a + S_b .
\label{13}
\een
Here, $ i S_F(x_1-x_2)$ is the fermion propagator (\ref{C6}).
Feynmann technology provides the formula
\eqn
\langle f | S^{(2)}|i \rangle = (2\pi )^3 \delta^{(3)} (p'+ k' - p - k) \prod
_{\rm ext} \left( {m \over A E_{\vec p}} \right) ^{1/2} \prod_{\rm ext} \left(
{1\over 2 A \omega_{\vec k} } \right) ^{1/2} ({\cal M}_a + {\cal M}_b)
\label{14}
\een
for the S-matrix element up to second order in perturbation, where the
Feynman amplitudes are
\beq
{\cal M}_a &=& - e^2 {\bar u}(\vec {p'}) \gamma^{\alpha} \epsilon_{\alpha}(\vec
{k'}) i S_F ( p + k) \gamma^{\beta} \epsilon_{\beta}(\vec k) u(\vec p) \nm \\
{\cal M}_b &=& - e^2 {\bar u}(\vec {p'}) \gamma^{\alpha} \epsilon_{\alpha}(\vec
{k }) i S_F ( p - k') \gamma^{\beta} \epsilon_{\beta}(\vec {k'}) u(\vec p) .
\label{15}
\eeq

The differential {\it cross-length} for this process is therefore
\eqn
d\lambda = (2\pi )^3 \delta^{(3)} (p' + k' - p - k) {(2 m)^2 \over 4 E \omega
v_{\rm rel} } \ {d^2 \vec {p'} \over (2\pi )^2 2 E' } \ {d^2 \vec {k'} \over
(2\pi )^2 2 \omega' }|{\cal M}|^2
\label{16}
\een
where $p=(E, \vec p)$ and $k = (\omega, \vec k)$ are the three-momenta for the
initial electron and photon, and the corresponding quantities for the final
electron and photon are $p'=(E', \vec {p'})$ and $k'=(\omega', \vec{k'})$.

Analysis of  the scattering of photons by electrons is easier in the
laboratory frame, in which $ p = (m, 0, 0)$ and $\vec {p'} = \vec k - \vec
{k'}$. The relative velocity in this system is unity, i. e., $v_{\rm
rel} =
{|\vec k| \over \omega} = 1$.

From the energy-momentum consevation law $\ p + k = p' + k' $, the Compton
shift in wavelength for
this process is easily deduced. In the laboratory system: $ p = (m, 0 ,0)$,
$ \vec k \cdot \vec {k'} = \omega \omega' \cos \theta$, $\theta$ is the
scattering angle,  and

\eqn
\omega' = { m \omega  \over m + \omega (1 - \cos \theta) } \label{17}
\een

There are no differences with the three-dimensional case
in this respect.
The recoil energy of the electron is
\eqn
E' = \sqrt{ m^2 + \omega^2 + {\omega'}^2 - 2 \omega \omega' \cos \theta}
\label{18}
\een

In (\ref{16}) we can integrate with respect to the dependent variables
$\vec {k'}$ and $\vec {p'}$
as a consecuence of the conservation of the initial and final momenta. Using
(\ref{17}) and (\ref{18}), we obtain the differential {\it cross-length} in the
laboratory frame
\eqn
\left( {d \lambda \over d \theta} \right) _{\rm Lab} = { 1 \over 8 \pi \omega}\
\left( {\omega' \over \omega} \right) \ |{\cal M}|^2 \label{19}
\een

In ${\rm QED}_{3+1}$ to obtain unpolarized cross-sections we must average
$|{\cal M}|^2$
over all the pola- rizations and spins of the initial state and sum it over
all final
polarizations and spin states. By doing this, we render the square of the
Feynman amplitudes
as the trace of products of $\gamma$-matrices. It is shown in Appendix
\ref{apdE} that despite
the lack of polarization or spin degrees of freedom in ${\rm QED}_{2+1}$
the square of the
Feynman amplitudes is also the trace of products of $\gamma$-matrices. The
right-hand member
of (\ref{19}), $|{\cal M}|^2$, is the sum of four terms:
\[
X_{aa} = {e^4 \over 16\, m^2 (pk)^2 } {\rm Tr} \left[ \gamma^{\beta} (
\gamma^{\mu} (p + k)_{\mu} + m)
\gamma^{\alpha} ( \gamma^{\nu} p_{\nu} + m) \gamma_{\alpha} (\gamma^{\rho}
(p + k)_{\rho} + m)
\gamma_{\beta} ( \gamma^{\lambda} {p'}_{\lambda} + m) \right]
\]
\eqn
\label{20}
\een
\[
 X_{ab} = {- e^4 \over 16  m^2 (pk) (pk') } {\rm Tr} \left[ \gamma^{\beta} (
\gamma^{\mu} (p+k)_{\mu}+m) \gamma^{\alpha} ( \gamma^{\nu} p_{\nu}+m)
\gamma_{\beta} (\gamma^{\rho}
(p-k')_{\rho}+m) \gamma_{\alpha} ( \gamma^{\lambda} {p'}_{\lambda}+m) \right]
\]
and $X_{bb} = X_{aa}( k \leftrightarrow - k', \epsilon \leftrightarrow
\epsilon')$, $X_{ba} = X_{ab}( k \leftrightarrow - k', \epsilon \leftrightarrow
\epsilon')$. Computation of the traces in (\ref{20}) is considerably
simplified by
the use of the contraction identities, see Appendix \ref{apdA}, because it
involves products
of up to eight $\gamma$-matrices. Note that the contraction identities for
$2\times 2$ $\gamma$-matrices are very different than the usual in
$(3+1)$-dimensions. In short, in terms of the three linearly independent scalars
$ p^2 = {p'}^2 = m^2 $, $ p k = p' k'$ and $p k' = p' k$ we have
\beq
&& X_{aa} = {e^4 \over 4\, m^2 (pk)^2 } \left[ 4 m^4 + 4 m^2 (pk) + (pk) (pk')
 \right]  \nm \\
&& X_{bb} = {e^4 \over 4\, m^2 (pk')^2 } \left[ 4 m^4 - 4 m^2 (pk') + (pk)
(pk')
 \right]  \label{21} \\
&& X_{ab} = -{e^4 \over 4\, m^2 (pk) (pk') } \left[  4 m^4 + 2 m^2 (pk - pk') -
(pk)(pk') + i 6 m \epsilon^{\mu \nu \lambda} p_{\mu} k_{\nu} k'_{\lambda}
\right]
\nm \\
&& X_{ba} = -{e^4 \over 4\, m^2 (pk) (pk') } \left[  4 m^4 + 2 m^2 (pk - pk') -
(pk)(pk') - i 6 m \epsilon^{\mu \nu \lambda} p_{\mu} k_{\nu} k'_{\lambda}
\right]
\nm
\eeq

In the laboratory system $pk = m \omega$, $pk' = m \omega'$ and from
(\ref{21}) and (\ref{19}) we obtain
the differential {\it cross-length} for the Compton scattering in the plane
\eqn
\left( {d \lambda \over d \theta } \right) _{\rm Lab} = {{\alpha}^2 \pi
\over 2 \omega} \ \left( {\omega' \over \omega} \right) \  \left\{ {\omega \over
\omega'} + {\omega' \over \omega} + 4 \cos^2 \theta - 2
\right\}      ;\label{22}
\een
 $  \left( {d \lambda \over d \theta } \right)$ has dimensions of
length (or $M^{-1}$ in the natural units system).

Plugging (\ref{17}) into (\ref{22}) and integrating the resulting
equation over the scattering angle, we find the total {\it cross-length}
\eqn
\lambda_{\rm total} = {\pi^2 \alpha^2 \over m \gamma } \left\{ {(1 + \gamma
) \over (1 + 2
\gamma )^{3/2} } + 1 + {4 (1 + \gamma) ( 1 + \gamma - \sqrt{1 + 2 \gamma })
\over \gamma^2 \sqrt{1 + 2 \gamma}}  -
{2 \over \sqrt{1 + 2 \gamma }} \right\}
\label{23}
\een
where $\gamma$ denotes the ratio of the photon initial energy to the
electron rest energy, i.e.,
$\gamma={\omega\over m}$ in natural units.

\subsection{Planar Thomson and Klein-Nishina formulas}

Contact between the experimental outcome of the Compton effect in the real
world and the theory is established
through the Klein-Nishina and Thomson formulas derived in ${\rm
QED}_{3+1}$. Also, the total cross-section offers
a direct connection between theory and experiment, which is particularly
fruitful at the non-relativistic and
extreme relativistic limits. Very good information about the behaviour of
photons when scattered by electrons
can be obtained by studying the angular distribution of the unpolarized
differential cross-section. In this
sub-section we discuss the same aspects in ${\rm QED}_{2+1}$ and compare
the results of the analysis with their
higher-dimensional counterparts.

Starting with the famous Klein-Nishina formula for the polarized
differential cross-section of Compton scattering:
\eqn
\left( {d \sigma \over d \Omega } \right) _{\rm Lab, pol} = { {\alpha}^2
\over 4 m^2} \ \left( {\omega' \over \omega} \right) ^2 \  \left\{ {\omega
\over \omega'} + {\omega' \over \omega} + 4 (\epsilon^{(\alpha)}
{\epsilon'}^{(\alpha')})^2 - 2 \right\}   ;
\label{24}
\een
we focus on the same magnitude in ${\rm QED}_{2+1}$. Before, however, let
us  notice that $\epsilon^{(\alpha)}\equiv
\epsilon^{(\alpha)}(\vec k)$ and ${\epsilon'}^{(\alpha')}\equiv
{\epsilon'}^{(\alpha')}(\vec k')$ are the
polarizations of the incident and scattered photons and $\alpha, \alpha' =
1, 2$.

It is convenient to write (\ref{24}) in the form
\eqn
\left( {d \sigma \over d \Omega } \right) _{\rm Lab, pol} = { {\alpha}^2
\over 4 m^2} \ \left( {\omega' \over \omega} \right) ^2 \  \left\{ {\omega
\over \omega'} + {\omega' \over \omega} + 4 \cos^2 \Theta - 2 \right\}   ;
\label{25}
\een
where $(\epsilon^{(\alpha)} {\epsilon'}^{(\alpha')})^2 = \cos^2 \Theta$ and
$\Theta$ is the angle formed by the
polarization vectors of the incident and scattered photons. In ${\rm
QED}_{2+1}$, the polarized differential
{\it cross-length} is given precisely by equation (\ref{22}). We therefore
call this expression the planar Klein-Nishina
formula; choosing the polarization vectors of the incident, $\vec
{\epsilon}^{(1)} (\vec k)$, and scattered,
$\vec{\epsilon'}^{(1)} (\vec k')$, photons in a
reference frame where the wave vector
$\vec k$ and $\vec {\epsilon}^{(1)} (\vec k)$ are respectively taken along
the z- and x-axes, the formula reads:
\eqn
\left( {d \lambda \over d \theta } \right) _{\rm Lab} = {{\alpha}^2 \pi
\over 2 \omega} \ \left( {\omega' \over \omega} \right) \  \left\{ {\omega \over
\omega'} + {\omega' \over \omega} + 4 (\epsilon^{(1)} {\epsilon'}^{(1)} )^2
- 2 \right\}. \label{26}
\een

The angle between the polarization vectors and the scattering angle now
coincide  \break $(  { \epsilon}^{(1)} \cdot
{\epsilon'}^{(1)} )^2 = \cos^2 \theta $. It is remarkable that one could
have derived the planar from the spatial Klein
-Nishina formula: taking the first of the polarization vectors
$\vec{\epsilon'}^{(1)} =(\cos \theta \cos \phi, \cos
\theta \sen \phi,$ $ - \sen \theta)$, we obtain: $\cos^2
\Theta = \cos^2 \theta \cos^2 \phi$. For
$\phi=0$ we almost recover the planar Klein-Nishina formula, but
dimensional reasons forbid a perfect identity between
both formulas and also differences between volume and area elements induce
some distinct factors. The other
polarization vector  $\vec{\epsilon'}^{(2)} =(\sen \phi, - \cos \phi, 0)$
is non-projectable to the plane, because
when $\phi=0$ it points in the direction which disappears.

A subtle point; in (2+1)-dimensions there is no difference between
polarized and unpolarized photon scattering
because planar photons have only one polarization. The differential {\it
cross-length} of scattering in ${\rm
QED}_{2+1}$ can also be compared with the unpolarized differential
cross-section in ${\rm QED}_{3+1}$. At the
non-relativistic limit it is given by the Thomson formula:
\eqn
\left( {d \sigma \over d \Omega} \right) _{\rm Lab, NR} = {r_0^2 \over 2}
(1 + \cos^2 \theta) \label{27}
\een
where $r_0 = {\alpha \over m}$ is the classical electron radius. At the NR
limit, where $\omega<<m$ and $ \omega'
\approx \omega$, we find from (\ref{22}) an analogous formula,
\eqn
\left( {d \lambda \over d \theta} \right) _{\rm Lab, NR} = {2 \pi \alpha^2
\over \omega} \cos^2 \theta \equiv {l_T
\over \pi \gamma} \cos^2 \theta \label{28}
\een
that we shall call the planar Thomson formula. Besides the scattering
angle, $\left( {d \lambda \over d \theta} \right)
_{\rm Lab, NR}$ depends on the parameter $\gamma =\omega/m$, the ratio of
the frequency of the incident photon to
the rest electron mass. Also, we have introduced a constant with dimensions
of length $l_T = (2 \pi^2 \alpha) r_0$
out of the two fundamental constants $\alpha$ and $r_0$, whose meaning for
the problem will be clear later. Unlike
the classical Thomson formula, the differential {\it cross-length} depends
on the incident photon energy at the
non-relativistic limit; in fact $\left( d \lambda / d \theta \right) _{\rm
Lab, NR}$ diverges when $\gamma
\rightarrow 0$. $\left( d \sigma / d \Omega \right) _{\rm Lab, NR}$,
however, is $\omega$-independent. The
intensity of the scattered radiation is thus higher when $\gamma$ decreases
in the planar Compton effect, but it
does not change when energy varies in (3+1)-dimensions. A common point is
that both differential scattering
cross-legths and cross-sections of the Compton effect are backward-forward
symmetric.

There are also noticeable differences in the behaviour of the scattering
differential {\it cross-length} and
unpolarized cross-section at the extreme relativistic limit $\omega >> m$.
If $\omega (1 - \cos \theta) << m$,
in this regime occuring at very small scattering angles, $\omega' \approx
\omega$ and:
\eqn
\theta \approx \delta \theta \  , \ \ \left( {d \lambda \over d \theta}
\right) _{\rm Lab,ER} \approx {l_T
\over \pi \gamma} \cos^2 \theta \  , \ \ \left( {d \sigma \over d \Omega}
\right) _{\rm Lab,ER} \approx {r_0^2
\over 2} (1 + \cos^2 \theta) \label{29}
\een

In this case, the intensity distribution of scattered radiation obeys the
same Thomson formulas as at the
non-relativistic limit. The {\it cross-lenght} at stake in the planar
Compton effect is smaller than the classical
one because the energy of the incoming photons is very high; in three
dimensions the cross-section at low energy and the
cross-section at high-energy and small enough angles are, however, the
same. If $\omega' = {m \over 1 - \cos
\theta}$ and $\omega (1- \cos \theta) >> m$, we are at the extreme relativistic limit looking at very large
scattering angles. Thus,
\eqn
\theta \approx {\pi \over 2} + \delta \theta \  , \ \  \left( {d \lambda
\over d \theta} \right) _{\rm Lab,ER}
\approx {l_T
\over 2 \pi \gamma} \  , \ \  \left( {d \sigma \over d \Omega} \right)
_{\rm Lab,ER} \approx {r_0^2
\over 2} {1\over \gamma (1 - \cos \theta)} \label{30}
\een
and the intensity of the planar Compton effect decreases with increasing
energy of the incoming photon, although it
is independent of the scattering angle. The latter feature is not shared by
the cross-section at the extreme
relativistic limit of ${\rm QED}_{3+1}$.

\begin{figure}[htbp]
\begin{center}
\epsfig{file=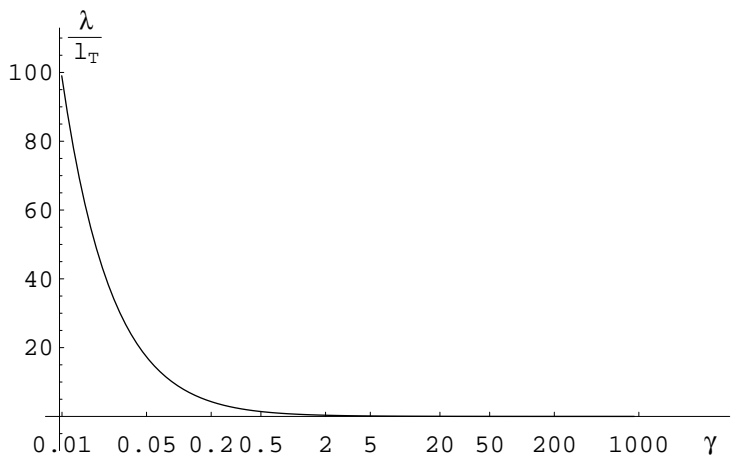,height=6cm}
\end{center}
\caption[Figure 1]{\small Total {\it cross-length} $\lambda/\l_T$ for the
planar Compton scattering as a function
of the initial photon energy $\gamma = w / m$ on a logarithmic scale. $l_T$
is the Thomson length
defined by $l_T =(2 \pi^2 \alpha) r_0$. }
\end{figure}

It is interesting to write the differential {\it cross-length} and the
unpolarized cross-section as functions of
$\gamma$:
\eqn
 \left( {d \lambda \over d \theta } \right) _{\rm Lab} = {l_T\over 4 \pi
\gamma} {1\over (1+ \gamma (1-\cos
\theta))}
\left\{ {1\over (1+ \gamma (1-\cos \theta))} + \gamma (1-\cos \theta) + 4
\cos^2 \theta - 1\right\} \label{31}
\een
\eqn
 \left( {d \sigma \over d \Omega } \right) _{\rm Lab} = {r_0^2 \over 2}
{1\over (1+ \gamma (1-\cos
\theta))^2}
\left\{ {1\over (1+ \gamma (1-\cos \theta))} + \gamma (1-\cos \theta) +
\cos^2 \theta  \right\} \label{32}
\een
From these expresions one computes respectively the total {\it cross-length}
\eqn
\lambda = {l_T \over 2  \gamma} \left\{ 1 + {1 + \gamma \over (1 + 2
\gamma)^{3/2} } + {4 (1 + \gamma) (1 +
\gamma - \sqrt{1 + 2 \gamma} ) - 2 \gamma^2 \over \gamma^2 \sqrt{1 + 2
\gamma} } \right\} \label{33}
\een
for the planar Compton effect, and the total cross-section
\eqn
\sigma = {2 \pi r_0^2}  \left\{ {1+ \gamma \over \gamma^3} \left( {2 \gamma
(1 + \gamma ) \over 1 + 2 \gamma }-
\log (1+ 2 \gamma) \right) + {1 \over 2 \gamma} \log(1 + 2 \gamma) - {1 + 3
\gamma \over (1 + 2 \gamma)^2}  \right\}
\label{34}
\een
for the same process in space \cite{Heitler}.

\begin{figure}[htbp]
\begin{center}
\epsfig{file=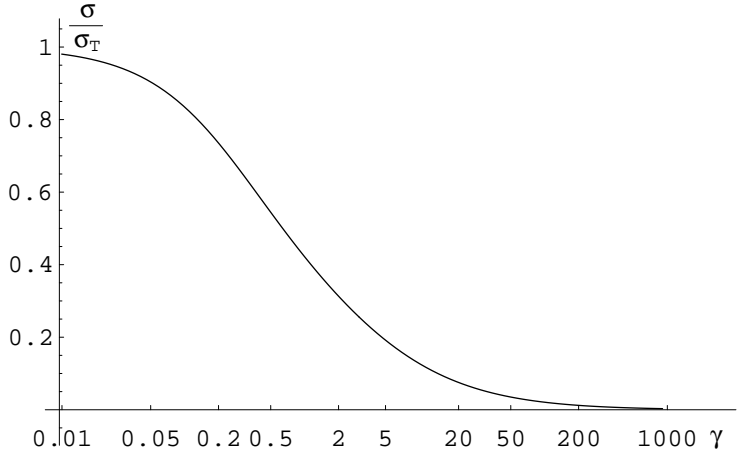,height=6cm}
\end{center}
\caption[Figure 2]{\small Total cross-section $\sigma/\sigma_T$ for Compton
scattering as a function of the
initial photon energy $\gamma = w / m$ on a logarithmic scale in order to
cover a large energy region. $\sigma_T$
is the cross-section for the Thomson scattering, \cite{Heitler}.}
\end{figure}

The non-relativistic limit, $\gamma << 1$, of $\lambda$ and $\sigma$ is
easily obtained
\beq
\lambda_{\rm NR} &=& {2 \pi^2 \alpha \over \gamma} r_0 \equiv {l_T \over
\gamma} \label{35} \\
\sigma_{\rm NR} &=& {8 \pi \over 3} r_0^2 = 6.65 \cdot 10^{-25} {\rm cm^2}.
\label{36}
\eeq
The cross-section $\sigma_T\equiv \sigma_{\rm NR}$ for Thomson scattering
is constant and independent of the incoming
photon frequency.
$\lambda_{\rm NR}$, however, depends on $\gamma$; we introduce a \lq \lq
natural" Thomson length $l_T = {2 \pi^2
\alpha^2 \over m} = 4.06\cdot 10^{-14} {\rm cm}$. It happens that for a
photon such that $\omega = 0.05 m = 0.0025
{\rm Mev}$, $\lambda_{\rm NR} = 8.12 \cdot 10^{-13} {\rm cm} \approx
\sqrt{\sigma_T}$. Other smaller values of
$\gamma$ lead to higher values of $\lambda_{\rm NR}$: for $\gamma = 0.03$
we find $\lambda_{\rm NR} = 1.35 \cdot
10^{-12} {\rm cm}$, for $\gamma=0.01$, $\lambda_{\rm NR} = 4.06 \cdot
10^{-12} {\rm cm}$, and so on.

At the other, extreme relativistic limit, $\gamma>>1$, we have
\beq
\lambda_{\rm ER} &=& {l_T \over 2} \left( {1\over \gamma} + {5 \over 2
\sqrt{2}} {1\over \gamma^{3/2}} \right)
\label{37} \\
\sigma_{\rm ER} &=& {3 \sigma_{T}\over 8}  \left( { 1 \over \gamma} \log
{2\gamma} + {1 \over 2 \gamma
} \right) \label{38}
\eeq
and we see that for very high energy  the Compton effect is a negligible
effect and pair
production becomes dominant both in the plane and in the three-dimensional
world. The logarithmic factor in
$\sigma_{\rm ER}$ announces that ultraviolet divergences in higher order
corrections will be more severe in
${\rm QED}_{3+1}$.

\begin{figure}[htbp]
\begin{center}
\epsfig{file=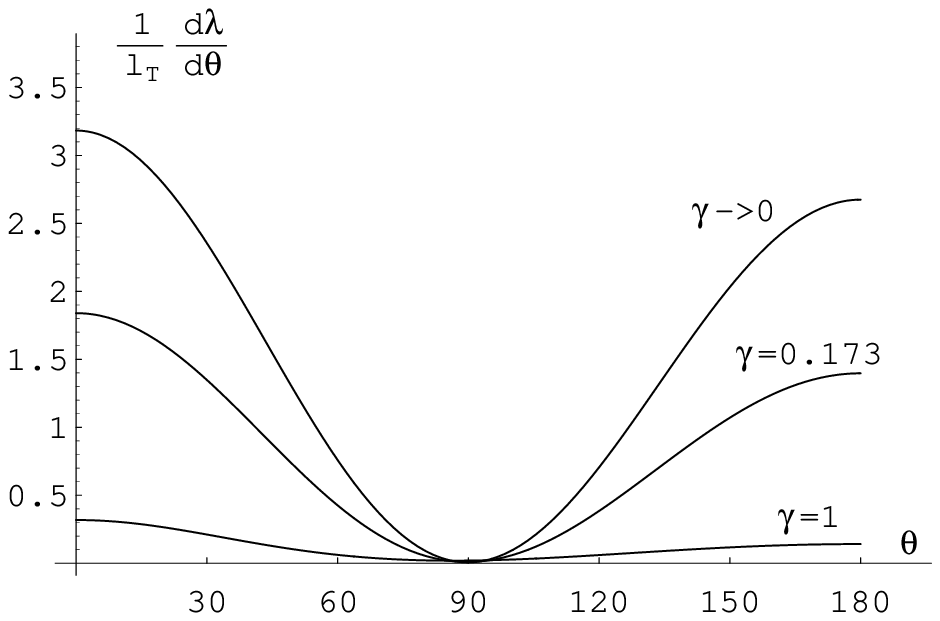,height=6cm}
\end{center}
\begin{center}
\epsfig{file=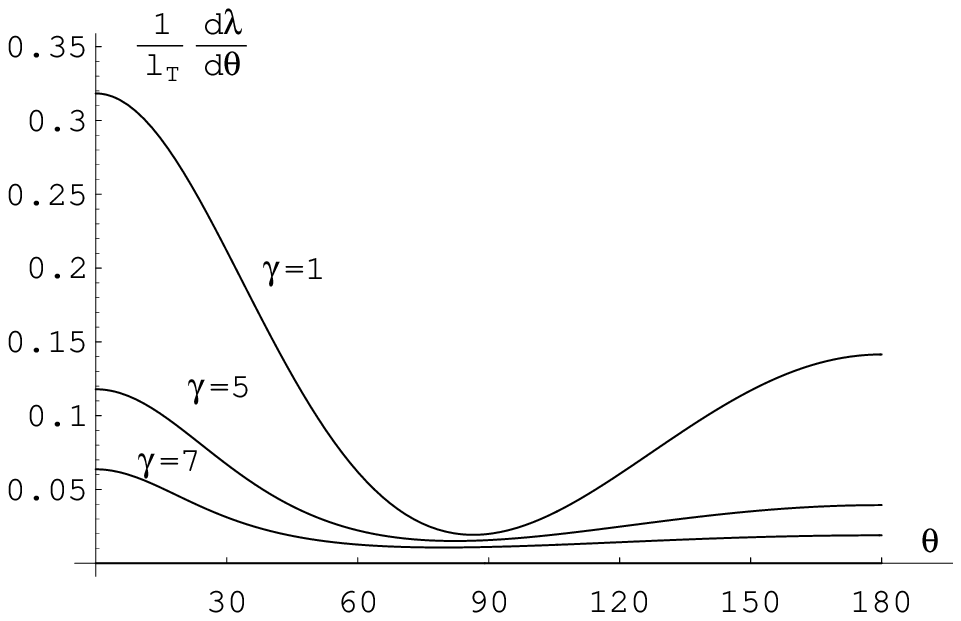,height=6cm}
\end{center}
\caption[Figure 3]{\small Angular distribution of the Planar Compton
scattering as a function of the
scattering angle  $\theta$ for several values of the initial photon energy
$\gamma = {\omega \over m}$.}
\end{figure}

The total {\it cross-length} $\lambda$ and cross-section $\sigma $ are
respectively plotted as functions of
$\gamma$ in Figures 1 and 2. The intensity of the scattered radiation is
very large for small energies in the
plane whereas in  space it is practically constant and equal to the
classical value. For high energy of the
incoming photons the intensity is very small in both cases, although it
goes to zero faster than in the first case.

Finally, we study how $\left( {d \lambda \over d \theta } \right) _{\rm
Lab}$ and $\left( {d \sigma \over d \Omega
} \right) _{\rm Lab}$ depend on the scattering angle. The angular
distribution of the differential
{\it cross-length} is plotted in Figure 3 for several chosen values of
$\gamma$ between the non-relativistic limit
and the high energy regime. A similar picture of $\left( {d \sigma \over d
\Omega
} \right) _{\rm Lab}$ is drawn in Figure 4. In both cases, the angular
distribution is forward-backward symmetric in
the non-relativistic limit $(\gamma\rightarrow 0)$, whereas in the
relativistic regime $(\gamma>>1)$ the forward
direction becomes more and more preponderant. For small angles, the
scattered intensity in the three-dimensional
case has almost the classical (non-relativistic) value for all the incident
energies; in the bi-dimensional case,
however, we observe that the higher the incident energy, the smaller the
intensity. Particularly, for
$\gamma\rightarrow 0$ the planar intensity is larger than the classical
value in Thomson scattering. For large
angles the angular distribution is similar in both cases but whereas in the
plane  $\left( {d \lambda \over d \theta
} \right) _{\rm Lab,NR}({\pi\over 2}) = 0$,  $\left( {d \sigma \over d
\Omega } \right) _{\rm Lab,NR}({\pi \over
2})$ is not zero. For high incident energies, the intensity is practically
equal to the minimum in the range $\theta
=(\pi/2, \pi)$.

A last comment on the pole found in $\left( {d \lambda \over d \theta }
\right) _{\rm Lab}$ at $\omega =0$: this
is an infrared divergence due to soft photons. This infrared catastrophe is
similar to that arising in
bremsstrahlung processes in ${\rm QED}_{3+1}$. Infrared divergences seem to
be more dangerous in ${\rm QED}_{2+1}$, but
fortunately, a
\lq \lq topological" mass for the photons is generated by the vacuum
polarization graph, see \cite{S.Deser82}, and a
natural infrared cut-off exists in the theory due to quantum corrections.

\begin{figure}[htbp]
\begin{center}
\epsfig{file=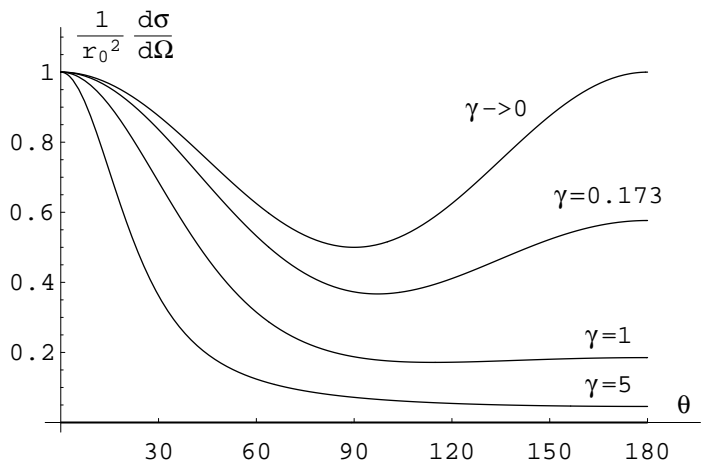,height=6cm}
\end{center}
\caption[Figure 4]{\small Angular distribution of the Compton scattering as
a function of the scattering angle
$\theta$ for several values of the initial photon energy $\gamma = {\omega
\over m}$ \cite{Heitler}.}
\end{figure}

\appendix

\section{Gamma Matrices in 3-dimensional Space-time}
\label{apdA}

The Dirac (Clifford) algebra in the 3-dimensional Minkowski space $M_3 =
{\bf R}^{1,2}$ is built from the
three gamma matrices $\gamma^{\mu}$ satisfying the anticommutation relations:
\eqn
\{ \gamma^{\mu}, \gamma^{\nu}\} = 2 g^{\mu \nu} \label{A1}
\een
\[
\mu = 0, 1, 2 \q , \q g^{\mu \nu} = {\rm diag} (1, -1, -1)
\]
and the hermiticity conditions  ${\gamma^{\mu}}^{\dag} = \gamma^0
\gamma^{\mu} \gamma^0$. The tensors
\eqn
1\,, \ \gamma^{\mu}\, , \ \gamma^{\mu_1} \gamma^{\mu_2}\, , \
\gamma^{\mu_1}\gamma^{\mu_2}\gamma^{\mu_3}\, ; \
\mu_1 < \mu_2 < \mu_3 \label{A2}
\een
with respect to the ${\rm SO}(2,1)$-group, the piece connected to the
identity of the Lorentz group in flatland,
form the basis of the Dirac algebra, which is thus $2^3$-dimensional. $1$ and
$\gamma^{\mu_1}\gamma^{\mu_2}\gamma^{\mu_3}= - i
\epsilon^{{\mu_1}{\mu_2}{\mu_3}} 1$ are respectively scalar and
pseudo-scalar objects. $\gamma^{\mu}$ is a three-vector but $\gamma^{\mu_1}
\gamma^{\mu_2}$ can be seen
alternatively as a anti-symmetric tensor or a pseudo-vector, which are
equivalent irreducible representations of
the ${\rm SO}(2,1)$-group. If we denote by $\epsilon^{\mu \nu \rho}$ the
completely antisymmetric tensor, equal to
+1(-1) for an even (odd) permutation of (0,1,2) and to 0 otherwise, the
$\gamma$-matrices must also satisfy
the commutation relations:
\eqn
\sigma^{\mu \nu} = {i \over 2}[ \gamma^{\mu}, \gamma^{\nu} ] =
\epsilon^{\mu \nu \rho} \gamma^{\rho}
\label{A3}
\een
The $\sigma^{\mu \nu}$-matrices are the Lie algebra generators of the $
{\rm spin}(1,2;{\rm \bf R}) \cong {\rm
SL}(2;{\rm \bf R})$-group, the universal covering of the connected piece of
the Lorentz group.

The irreducible representations of the Lie ${\rm SL}(2;{\rm \bf R})$-group
are the spinors and, before choosing
a particular representation suitable for describing planar electrons and
positrons, we list some algebraic identities
that are useful in the evaluation of the trace of $\gamma$-matrix products:
\begin{itemize}
\item $\gamma$-matrix contractions:
\beq
& & \gamma_{\mu} \gamma^{\mu} = 3  \q, \q \gamma_{\mu} \gamma^{\nu}
\gamma^{\lambda} \gamma^{\mu} = 4 g^{\nu
\lambda} - \gamma^{\nu} \gamma^{\lambda}  \label{A4} \\
& & \gamma_{\mu} \gamma^{\nu} \gamma^{\mu} = - \gamma^{\nu} \q, \q
 \gamma_{\mu} \gamma^{\nu} \gamma^{\lambda} \gamma^{\delta} \gamma^{\mu} = -
2 \gamma^{\delta} \gamma^{\lambda} \gamma^{\nu} + \gamma^{\nu} \gamma^{\lambda}
\gamma^{\delta} \nm
\eeq
\item $\epsilon$-products:
\eqn
\epsilon^{\alpha \beta \delta} \epsilon_{\alpha \lambda \mu} =
g_{\lambda}^{\beta} g_{\mu}^{\delta} - g_{\mu}^{\beta} g_{\lambda}^{\delta}
\q , \q
 \epsilon^{\alpha \beta \delta}
\epsilon_{\alpha \beta \mu} = 2 g_{\mu}^{\delta} \q , \q
 \epsilon^{\alpha \beta \delta} \epsilon_{\alpha \beta
\delta} = 6 \label{A5}
\een
\item Traces of products of $\gamma$-matrices:
\beq
& & {\rm Tr} (\gamma^{\mu}) = 0 = {\rm Tr} (\sigma^{\mu \nu})\q , \q
 {\rm Tr} (\gamma^{\mu} \gamma^{\nu}) = 2 g^{\mu \nu} \label{A6} \\
& & {\rm Tr} (\gamma^{\mu} \gamma^{\nu} \gamma^{\lambda}) = - 2 i \epsilon^{\mu
\nu \lambda} \ , \  {\rm Tr} (\gamma^{\mu} \gamma^{\nu}\gamma^{\lambda}
\gamma^{\delta}) = 2
\left(  g^{\mu \nu} g^{\lambda \delta} - g^{\mu \lambda} g^{\nu \delta} + g^{\mu
\delta} g^{\nu \lambda} \right) \nm
\eeq
\item For an even number of $\gamma$-matrices,
\eqn
{\rm Tr} (\gamma^{\mu} \gamma^{\nu} \cdots \gamma^{\delta}
\gamma^{\lambda}) = {\rm Tr} (\gamma^{\lambda} \gamma^{\delta} \cdots
\gamma^{\nu} \gamma^{\mu}) \label{A7}
\een
but this identity is false for an odd number of $\gamma$-matrices. For
instance, ${\rm Tr} (\gamma^{\mu}
\gamma^{\nu} \gamma^{\lambda}) = - {\rm Tr} (\gamma^{\lambda} \gamma^{\nu}
\gamma^{\mu})$.

\end{itemize}

\section{The Dirac Equation.}
\label{apdB}

Massive classical Dirac fields satisfy the momentum space Dirac equations
at rest:
\eqn
(\gamma^0 - 1) u(0) = 0 \q \q (\gamma^0 + 1 ) v(0) = 0 \label{B1}
\een
The spinors $u(0)$ and $v(0)$ are eigenfunctions of the spin matrix ${\hbar
\over 2} \sigma^{12} = {\hbar \over 2}
\gamma^0$ with $\pm {\hbar \over 2}$ eigenvalues. In a spinor
representation, a general Lorentz transformation, $p'_{\mu} =
\Lambda_{\mu}^{\nu} p_{\nu}$,
$p=({E_{\vec p}\over c}, \vec p)$ with $\vec p =(p_1,p_2)$ and $E_{\vec p}
= + \sqrt{ m^2 c^4 + c^2 {\vec p} {\vec
p} }
$, is given by:
\eqn
f'(p') = {\rm exp} \left\{ {i\over 2} \sigma^{\mu \nu} \omega_{\mu \nu}
\right\} f(\Lambda p) \label{B2}
\een
Lorentz boosts arise when $\omega_{0i}\neq 0$ and $\sigma^{12}$ is thus the
generator of rotations at the centre of
mass of the system. Applying a pure Lorentz transformation to (\ref{B1}),
we obtain the Dirac equations
\eqn
 ( \gamma^{\mu} p_{\mu} - m c) u(\vec p) = 0 \q  \q (\gamma^{\mu} p_{\mu} +
m c) v(\vec p) = 0 \label{B3}
\een
which are automatically Lorentz-invariant. We also write the conjugate
equations satisfied by the Dirac adjoint
spinors $\bar u(\vec p) = u^{\dag}(\vec p) \gamma^0 \, , \, \bar v(\vec p)
= v^{\dag}(\vec p) \gamma^0$:
\eqn
{\bar u} (\vec p)  (\gamma^{\mu} p_{\mu} - m c ) = 0  \q   \q {\bar v}
(\vec p)  (\gamma^{\mu} p_{\mu} + m c ) = 0
 \label{B4}
\een

Choosing the normalizations as follows,
\[
u^{\dag} (\vec p) u(\vec p) =  v^{\dag} (\vec p) v(\vec p) = {E_{\vec p} \over m
c^2} \q , \q {\bar u} (\vec p) u(\vec p) = - {\bar v} (\vec p) v(\vec p) = 1
\]
and given the orthogonality conditions,
\[
 u^{\dag} (\vec p) v(- \vec p) = 0 \q ,\q
{\bar u} (\vec p) v( \vec p) = {\bar v}(\vec p) u(\vec p) = 0
\]
these spinors satisfy the completeness relation:
\eqn
u_{\alpha}(\vec p) {\bar u}_{\beta} (\vec p) - v_{\alpha} (\vec p) {\bar
v}_{\beta}
(\vec p)  = \delta_{\alpha \beta} \ , \ \alpha, \beta = 1, 2 \label{B5}
\een
in a representation of the Dirac algebra of minimal dimension.

In fact, in 3-dimensional Minkowski space (pseudo)-Majorana spinors do
exist: in this representation of the Clifford
Algebra, the $\gamma$-matrices are purely imaginary $2\times 2$ matrices
and the spinors have two real components. We
wish to describe charged particles, so we choose the representation of the
Dirac algebra as:
\eqn
\gamma^0 = \sigma_3 \ , \ \gamma^1 = i \sigma_1 \ , \ \gamma^2 = i \sigma_2
\label{B6}
\een
where the $\sigma^a\, ,\, a = 1, 2, 3$ are the Pauli matrices. In this
representation the normalized solutions of
(\ref{B3}) are:
\eqn
 u(\vec p) = \sqrt{ E_{\vec p} + m c^2 \over 2 m c^2} \left( \ba{c} 1 \\ {c
(p_2-ip_1) \over E_{\vec p}+ m c^2} \ea \right) \q , \q
 v(\vec p) = \sqrt{ E_{\vec p} + m c^2 \over 2 m c^2} \left( \ba{c}  {c
(p_2+ip_1) \over E_{\vec p}+ m c^2} \\ 1 \ea \right) \label{B7}æ
\een

Putting the system in a normalization square of large but finite area, $A =
L^2$, we can expand the classical Dirac
field in a Fourier series:
\beq
\psi(x) &=& \psi^+ (x) + \psi^- (x) \nm \\
&=& \sum_{\vec p} \left( {m c^2 \over A E_{\vec p}} \right) ^{1/2}
\left[ c(\vec p) \, u(\vec p)\, e^{- {i p x \over \hbar}} + d^{*}(\vec p)\,
v(\vec p)\,
e^{i p x \over \hbar} \right] \label{B8} \\
\bar{\psi}(x) &=& {\bar \psi}^+ (x) + {\bar \psi}^- (x) \nm \\
&=& \sum_{\vec p} \left( {m c^2 \over A E_{\vec p}} \right) ^{1/2}
\left[ d(\vec p)\, {\bar u}(\vec p)\, e^{- {i p x \over \hbar}} +
c^{*}(\vec p)\,
{\bar v}(\vec p)\, e^{i p x \over \hbar} \right] \nm
\eeq
The Fourier transform of equations (\ref{B3}) and (\ref{B4}) are the Dirac
equation and its conjugate in
the configuration space:
\eqn
i \hbar \gamma^{\mu}{\partial \psi(x) \over \partial x^{\mu}} - m c \psi(x) =
0 \q , \q
i \hbar {\partial \bar{\psi}(x) \over \partial x^{\mu}}æ\gamma^{\mu} + m c
\bar{\psi} (x) = 0 \label{B9}
\een
The plane wave expansions (\ref{B8}), where $c(\vec p)$ and $d(\vec p)$ are
the Fourier coefficients, solve
(\ref{B9}).

The symmetry group of classical $({\rm CED})_{2+1}$ is the Poincar\'e
group, the semi-direct product of the Lorentz group times the
abelian group of translations in Minkowski space. The two sheets of the
hyperboloid $p^2 = m^2$,
\[
{\hat O}_m^+ = \{ p\in {\hat M}_3\, ,\, p^2 = m^2,\, p_0 > 0\}\ ,\  {\hat
O}_m^- = \{ p\in {\hat M}_3,\, p^2 =
m^2,\, p_0 < 0 \},
\]
are disconnected orbits of the Lorentz group in the dual of Minkowski space
${\hat M}_3$. Equations (\ref{B1}) hold
at the points $p = ( \pm m, 0, 0)$, respectively. Thus, solutions on ${\hat
O}_m$ correspond to $E_{\vec p} = \pm m c^2$
and, after quantization, $v(\vec p)$ will be interpreted as the antiparticle spinor. There is a very important
difference with the situation in 4-dimensional space-time: the isotropy
group of the $p = ( \pm m,0,0)$ points is
now ${\rm SO} (2)$, which has ${\bf R}$ as covering group. Therefore, the
spin is a scalar in (2+1)-dimensions having any real value because the
irreducible representations of an abelian group such as ${\rm \bf R}$ are
one-dimensional.

One can check that the Dirac equation is the Euler-Lagrange equation for
the Lagrangian:
\eqn
{\cal L} = c \bar{\psi}(x) \left( i \hbar \gamma^{\mu} {\partial \over \partial
x^{\mu}} - m c \right) \psi (x) \label{B10}
\een
which, besides the invariance with respect to the Poincar\'e group
transformations connected to the identity, is
invariant under the discrete transformations of parity and time-reversal:

\n  1)
\beq
&& {\vec x'} = ({x'}^1, {x'}^2) = (- x^1, x^2) \nm \\
&& P \psi(x^0, \vec x) P^{-1} = \sigma^1 \psi(x^0, {\vec x'}) \nm
\eeq
 \n 2)
\beq
&& { x'}^0 = - x^0 \nm \\
&& T \psi(x^0, \vec x) T^{-1} = \sigma^2 \psi({x'}^0, {\vec x}) \nm
\eeq

Finally, the energy projection operators $\Lambda^{\pm} (\vec p) = { \pm
\gamma^{\mu} p_{\mu} + m c \over 2 m c}$
are
\eqn
\Lambda^{+}_{\alpha \beta} (\vec p) = u_{\alpha}(\vec p) {\bar
u}_{\beta} (\vec p) \q , \q
\Lambda^{-}_{\alpha \beta} (\vec p) = - v_{\alpha} (\vec p) {\bar v}_{\beta}
(\vec p)  \label{B11}
\een

\section{ The Dirac Field}
\label{apdC}

From the Lagrangian, we obtain the Dirac Hamiltonian
\eqn
H_D = \int d^2 x \left[ \psi^{\dag}(x) \left( c {\vec \alpha} \cdot ( - i
\hbar {\vec
\nabla}) + \beta m c^2 \right) \psi(x) \right] \label{C1}
\een
where $\beta = \gamma^0$ and $ \alpha^j = \beta \gamma^j$, $j= 1, 2$, and
quantize the system by promoting the Fourier
coefficients to quantum operators which satisfy the anticommutation relations:
\eqn
\{ c(\vec p) \ , \ c^{\dag}(\vec{ p'}) \} = \{ d(\vec p) \ , \ d^{\dag}(\vec{
p'}) \} = \delta_{{\vec p}, \vec{ p'}} \label{C2}
\een
and all other anticommutators vanish. $c$ and $c^{\dag}$ are the
annihilation and creation operators of electrons,
while $d$ and $d^{\dag}$ play a similar role with respect to planar
positrons. The fermionic Fock space is built
out of the vacuum,
\[
c(\vec p) | 0 \rangle = d(\vec p) | 0 \rangle = 0 \q , \q \forall \vec p
\]
by the action of strings of creation operators:
\[
|n({\vec p}_1^+)n({\vec p}_1^-) \cdots n({\vec p}_N^+)n({\vec p}_N^+)
\rangle \propto [d^{\dag}({\vec p}_1^+)]^{
n({\vec p}_1^+)}[c^{\dag}({\vec p}_1^-)]^{
n({\vec p}_1^-)}\cdots [d^{\dag}({\vec p}_N^+)]^{
n({\vec p}_N^+)}[c^{\dag}({\vec p}_N^+)]^{
n({\vec p}_N^+)} |0 \rangle
\]
where $n({\vec p}_a^{\pm}) = 0$ or 1 due to the Fermi statistics coming
from (\ref{C2}).

From (\ref{C2}) and the plane wave expansion (\ref{B8}) one obtains
\eqn
 \{ \psi(x), \psi(y) \} = \{ \bar{\psi}(x), \bar{\psi}(y) \} = 0 \q , \q
 \{ \psi_{\alpha} (x), \bar{\psi}_{\beta}(y) \} = i S_{\alpha \beta} (x -y)
\label{C3}
\een
where the $2\times 2$-matrix function $S(x) = S^+ (x) + S^- (x)$ is given by
\eqn
S^{\pm} (x) =  \left( i \gamma^{\mu} {\partial \over \partial x^{\mu}} + {m
c \over
\hbar} \right)  \Delta^{\pm} (x) \label{C4}
\een
Here, $\Delta^{\pm} (x)$ are the invariant $\Delta$-functions, see
\cite{Mandl}, that admit
the integral representation
\eqn
 S^{\pm} (x) = {- \hbar \over (2 \pi \hbar)^3 } \int_{C^{\pm}} d^3 p \
e^{-{i p x
\over \hbar}} { \gamma^{\mu} p_{\mu} + m c \over p^2 - m^2 c^2} \label{C5}
\een
if $C^{\pm}$ are the contours in the complex $p_0$-plane that enclose the poles
at $p_0 = \pm (E_{\vec p} /c)$.

The fermion propagator $S_F(x-y)$ is the expectation value of the
time-ordered product $ T\{ \psi(x) {\bar \psi}
(y)\}$ at the vacuum state:
\beq
 i S_F (x-y) &=& \langle 0 | T\{ \psi(x) {\bar \psi} (y) \}  | 0\rangle =
\nm \\
&= & i ( \theta(x^0 - y^0) S^+ (x-y) - \theta (y^0-x^0) S^- (x-y)) =  \\
&=& { i\hbar \over (2 \pi \hbar)^3} \int d^3 p \ e^{- {i p (x-y) \over
\hbar}} { \gamma^{\mu}
p_{\mu} + m c \over p^2 - m^2 c^2 + i \epsilon } \label{C6}
\eeq
$\theta(x)$ is the step function, $\theta(x) = 1$ if $x>0$, $\theta(x) = 0$ if $x >0$.

The Dirac field at $\vec p = 0$
\eqn
\psi(x) = {c \over L} \left[ c(0) \left( \ba{c} 1 \\ 0 \ea \right) e^{- i{m
c^2 \over \hbar } t} + d^{\dag} (0)
\left( \ba{c} 0 \\ 1 \ea \right) e^{i {m c^2 \over \hbar }t} \right] \label{C7}
\een
destroys an electron of spin $\hbar /2$ and creates one positron of spin $
\hbar/2$. Quanta with any $\vec p$ are
obtained from the center of mass states by the action of Lorentz boosts.
Unlike in four-dimensional space-time,
we cannot talk of helicity in a purely three-dimensional universe because
the spin is a scalar.

\section{ The electromagnetic field in $(2+1)$-dimensions.}
\label{apdD}

The canonical quantization of the electromagnetic field in $(2+1)$-dimensions is
equivalent to the four-dimensional case. We shall follow the covariant
formalism
of Gupta and Bleuler, see \cite{Mandl}. We consider the Fermi Lagrangian
density
\eqn
{\cal L} = -{1\over 2} \left( \partial_{\nu} a_{\mu}(x) \right) \left(
\partial^{\nu} a^{\mu}(x) \right) \label{D1}
\een
where now $a^{\mu}(x) \, , \, \mu = 0, 1, 2$ is the three-vector potential. The
fields equations are
\eqn
\Box a^{\mu}(x) = 0 \label{D2}
\een
which are equivalent to Maxwell's equations if the potential satisfies the
Lorentz condition $\partial_{\mu} a^{\mu} (x) = 0$. We expand the free
electromagnetic field in a complete set of plane wave states:
\beq
a^{\mu}(x) &=& a^{\mu +}(x) + a^{\mu -}(x)  \nm \\
& = & \sum_{\vec k, r} {1 \over \sqrt{2 A \omega_{\vec k}} }
\left( \epsilon_r^{\mu}(\vec k)\, b_r(\vec k)\, e^{-i k x} +
\epsilon_r^{\mu}(\vec k)\,
b_r^{\dag} (\vec k)\, e^{i k x} \right) \label{D3}
\eeq
Here, the summation is over wave vectors, allowed by the periodic boundary
conditions in $A$, with $k^0 = {1 \over c} \omega_{\vec k} = |
\vec k|$. The summation over $r = 0, 1, 2$ corresponds to the three linearly
independent polarizations states that exist for each $\vec k$. The real
polarization vectors $\epsilon_r^{\mu} (\vec k)$ satisfy the orthonormality and
completeness relations
\beq
 & & \epsilon_{r \mu} (\vec k) \epsilon_s^{\mu}(\vec k) = - \eta_r
\delta_{rs}, \q
r, s = 0, 1, 2 \label{D4} \\
& & \sum_{r} \eta_r \epsilon_r^{\mu}(\vec k) \epsilon_r^{\nu}(\vec k) = - g^{\mu
\nu} \label{D5} \\
& &  \eta_0 = -1 \ ,  \ \eta_1 = \eta_2 = 1 \nm
\eeq

The equal-time commutation relations for the fields $a^{\mu}(x)$ and their
momenta $\pi^{\mu}(x) = -{1\over c^2} {\dot a}^{\mu}(x)$ are
\beq
& & [ a^{\mu} (\vec x, t), a^{\nu} (\vec {x'}, t) ] =  [ {\dot a}^{\mu} (\vec x,
t), {\dot a}^{\nu} (\vec {x'}, t) ] = 0 \nm \\
& & [ a^{\mu} (\vec x, t), {\dot a}^{\nu} (\vec {x'}, t) ] = - i \hbar c^2
g^{\mu
\nu} \delta^{(2)} (\vec x - \vec {x'}) \label{D6}
\eeq
The operators $b_r(\vec k) $ and $b_r^{\dag}(\vec k)$ satisfy
\eqn
[ b_r(\vec k), b_s^{\dag} (\vec {k'}) ] = \eta_r \delta_{r s} \delta_{\vec k
\vec {k'}} \label{D7}
\een
and all other commutators vanish. For each value of $r$ there are transverse $(r
=1)$, longitudinal $(r=2)$ and scalar $(r=0)$ photons, but as result of the
Lorentz
condition, which in the Gupta-Bleuler theory is replaced by a restriction on the
states, only transverse photons are observed as free particles. This is
accomplished as follows: the states of the
basis of the bosonic Fock space have the form,
\[
|n_{r_1}(\vec{k_1}) n_{r_2}(\vec{k_2}) \cdots n_{r_N}(\vec{k_N}) \rangle
\propto \left[ a_{r_1}^{\dag}
(\vec{k_1})\right] ^{n_{r_1}(\vec{k_1})}  \left[ a_{r_2}^{\dag}
(\vec{k_2})\right] ^{n_{r_2}(\vec{k_2})} \cdots  \left[ a_{r_N}^{\dag}
(\vec{k_N})\right] ^{n_{r_N}(\vec{k_N})} |0 \rangle ,
\]
where $ n_{r_i}(\vec{k_i}) \in {\bf Z}^+$, $\forall i = 1, 2, \cdots, N$ and
\[
a_r(\vec k) |0 \rangle = 0 \ , \ r = 0, 1, 2
\]
defines the vacuum state. To avoid negative norm states the condition
\[
\left[ a_2(\vec k)  - a_0 (\vec k) \right] | \Psi \rangle = 0 \ , \
\forall \ \vec k \Longleftrightarrow
\langle \Psi |N_2(\vec k) |\Psi \rangle = \langle \Psi |N_0 (\vec k) |\Psi
\rangle
\]
is required on the physical photon states of the Hilbert space. Therefore,
in two dimensions, there is only one
degree of freedom for each $\vec k$ of the radiation field.

From the covariant commutation relations we derive the Feynman photon
propagator:
\eqn
\langle 0 | T \{ a^{\mu} (x) a^{\nu} (y) \} | 0 \rangle = i \hbar c
D_F^{\mu \nu}
(x - y) \label{D8}
\een
where
\eqn
D_F^{\mu \nu}(x) = {1 \over (2 \pi )^3 } \int d^3 k {- g^{\mu \nu} \over k^2 + i
\epsilon} e^{-i k x} \label{D9}
\een

Choosing the polarization vectors in a given frame of reference as
\beq
& & \epsilon_0^{\mu}(\vec k) = n^{\mu} = (1, 0, 0) \nm \\
& & \epsilon_1^{\mu}(\vec k) = (0, {\vec \epsilon}_1(\vec k) ) \ , \ {\vec
\epsilon}_1(\vec k)  \cdot \vec k = 0 \label{D10} \\
& & \epsilon_2^{\mu}(\vec k) = (0, {\vec k \over |\vec k|}) = {k^{\mu} - (k
n) n^{\mu} \over ( (kn)^2 -
k^2)^{1/2}}  \nm
\eeq
it is possible to express the momentum space propagator from (\ref{D9})
as
\beq
D_F^{\mu \nu}(k) &=& {- g^{\mu \nu} \over k^2 + i \epsilon} \nm \\
&=& D_{FT}^{\muæ\nu}(k) + D_{FC}^{\mu \nu}(k) + D_{FR}^{\mu \nu}(k)
\label{D11} \\
&=&  {1 \over k^2 + i \epsilon} \epsilon_1^{\mu} (\vec k)
\epsilon_1^{\nu}(\vec k)
+  {n^{\mu} n^{\nu} \over (kn)^2 - k^2}  + {1\over k^2 + i \epsilon} \left[
{k^{\mu} k^{\nu} -(kn) (k^{\mu} n^{\nu} + k^{\nu} n^{\mu}) \over (kn)^2 -k^2}
\right] \nm
\eeq

The first term in (\ref{D11}) can be interpreted as the exchange of transverse
photons. The remaining two terms follow from a linear combination of
longitudinal and temporal photons such that
\eqn
D_{FC}^{\mu \nu}(x) = {g^{\mu 0} g^{\nu 0} \over (2 \pi)^3 }\int {d^2 \vec
k \ \ e^{i
\vec k \cdot \vec x} \over |\vec k|^2 } \int d k^0 e^{- i k^0 x^0} =  g^{\mu 0}
g^{\nu 0} {1\over 4 \pi } \ln {1\over |\vec x|} \delta( x^0)    ;\label{D12}
\een
This term corresponds to the instantaneous Coulomb interaction between
charges in the
plane, and the contribution of the remainding term $D_{FR}^{\mu \nu}(k)$
vanishes
because the electromagnetic field only interacts with the conseved
charge-current density, \cite{Mandl}.

\section{ Spin and polarization sums in $(2+1)$ dimensions.}
\label{apdE}

In ${\rm QED}_{3+1}$ the unpolarized cross-section is obtained by
averaging  $|{\cal M}|^2$ over all initial polarization states and summing
it over all
final polarization states. However, in ${\rm QED}_{2+1}$ we have seen on the one
hand that there are not degrees of freedom concerning the spin of the particles
and, on the other hand, that there is only one transverse polarization
state for the
photon: in the plane there is no need to average  and summ over all the
polarization
states. However, it is possible to show that $|{\cal M}|^2$ can be written
in one of the two forms:
\begin{itemize}
\item  1. We consider a Feynman amplitude ${\cal M} = {\bar u}(\vec {p'})
\Gamma u(\vec p) $, where $ u (\vec {p})$ and ${\bar u}(\vec {p'})$ are two
component spinors that specify the momenta of the electron in the initial and
final states, and $\Gamma$ is a $2\times 2$ matrix built up out of
$\gamma$-matrices. Then
\beq
|{\cal M}|^2 &=& {\bar u}(\vec {p'}) \Gamma u(\vec p) {\bar u}(\vec {p})
\tilde{ \Gamma} u(\vec {p'}) \nm \\
&=& \left( u_{\alpha}(\vec {p'}) {\bar u}_{\beta}(\vec {p'})\right)
\Gamma_{\beta \gamma} \left( u_{\gamma}(\vec
p) {\bar u}_{\delta}(\vec {p}) \right) \tilde{ \Gamma}_{\delta \alpha} \nm \\
&=& \Delta_{\alpha \beta}^+ (\vec p') \Gamma_{\beta \gamma} \Delta_{\gamma
\delta}^+(\vec p) {\tilde
\Gamma}_{\delta \alpha} \nm \\
&=& {\rm Tr} \left[ {\gamma^{\mu} {p'}_{\mu} + m \over 2 m} \Gamma
{\gamma^{\nu} p_{\nu} + m
\over 2 m} \tilde{\Gamma} \right] \label{E1}
\eeq
where we have used the positive energy projection operator (\ref{B11}) and
$\tilde \Gamma = \gamma^0 \Gamma^{\dag}
\gamma^0$. This can be extended to Feynman amplitudes with one or two
spinors of antiparticles using the negative
energy projection operator (\ref{B11}).

\item 2. We consider a Feynman amplitude of the form ${\cal M} =
\epsilon_1^{\alpha}(\vec k) {\cal M}_{\alpha}(\vec k) $, i. e., with one
external
photon. The gauge invariance requires $k^{\alpha} {\cal M}_{\alpha}(\vec k) = 0$
so that:
\eqn
|{\cal M}|^2 = \epsilon_1^{\alpha}(\vec k) \epsilon_1^{\beta}(\vec k)
{\cal M}_{\alpha}(\vec k) {\cal M}_{\beta}^{*}(\vec k) = -{\cal
M}^{\alpha}(\vec k)
{\cal M}_{\alpha}^{*} (\vec k)    ;\label{E2}
\een
Here we have used the relation
\eqn
\epsilon_1^{\alpha}(\vec k) \epsilon_1^{\beta}(\vec k)  = - g^{\alpha \beta }
- \left( {k^{\alpha} k^{\beta} - (kn)(k^{\alpha} n^{\beta} + n^{\alpha}
k^{\beta}) \over (kn)^2 } \right) \label{E3}
\een
for a physical  photon $k^2 =0$. Once again, this formalism can be extended
to several
external photons.

\end{itemize}

\end{document}